\documentclass{egpubl}
\usepackage{eurovis2020}

\SpecialIssuePaper         

\usepackage[T1]{fontenc}
\usepackage{dfadobe}

\usepackage{cite}  
\BibtexOrBiblatex
\electronicVersion
\PrintedOrElectronic
\ifpdf \usepackage[pdftex]{graphicx} \pdfcompresslevel=9
\else \usepackage[dvips]{graphicx} \fi

\usepackage{egweblnk}

\usepackage[bb=boondox]{mathalfa}  

\usepackage[normalem]{ulem}
\title[Quantitative Evaluation of Time-Dependent Multidimensional Projection Techniques]%
      {Quantitative Evaluation of Time-Dependent Multidimensional Projection Techniques}

\author[Vernier \textit{et al.}]
{\parbox{\textwidth}{\centering
E.\,F. Vernier$^{1,2}$,
R. Garcia$^{1}$,
I.\,P. da Silva$^{1}$,
J.\,L.\,D. Comba$^{1}$ and
A.\,C. Telea$^{3}$
        }
        \\
{\parbox{\textwidth}{\centering $^1$Federal University of Rio Grande do Sul, Brazil\\
         $^2$University of Groningen, the Netherlands\\
         $^3$University of Utrecht, the Netherlands
       }
}
}

\begin{document}

\renewcommand{\floatpagefraction}{.9}%


\maketitle
\begin{abstract}
Dimensionality reduction methods are an essential tool for multidimensional data analysis, and many interesting processes can be studied as time-dependent multivariate datasets. There are, however, few studies and proposals that leverage on the concise power of expression of projections in the context of dynamic/temporal data.  In this paper, we aim at providing an approach to assess projection techniques for dynamic data and understand the relationship between visual quality and stability. Our approach relies on an experimental setup that consists of existing techniques designed for time-dependent data and new variations of static methods. To support the evaluation of these techniques, we provide a collection of datasets that has a wide variety of traits that encode dynamic patterns, as well as a set of spatial and temporal stability metrics that assess the quality of the layouts.  We present an evaluation of 11 methods, 10 datasets, and 12 quality metrics, and elect the best-suited methods for projecting time-dependent multivariate data, exploring the design choices and characteristics of each method. All our results are documented and made available in a public repository to allow reproducibility of results.



\ccsdesc[2012]{Human-centered computing~Visualization techniques}
\ccsdesc[2012]{Computing methodologies~Dimensionality reduction and manifold learning}

\printccsdesc
\end{abstract}
\section{Introduction}
Dimensionality reduction (DR) methods, also called projections, are used in many applications in information visualization, machine learning, and statistics. Compared to other high-dimensional data visualization techniques, projections are especially effective for datasets with many observations (also called samples or points) and attributes (also called measurements, dimensions, or variables)\,\cite{Liu2017}. Many projection techniques exist, with wide varieties of computational efficiency, ease of use, ability to preserve and/or enhance different data patterns. Surveys have also focused on assessing quantitative and qualitative aspects of projection techniques\,\cite{nonato_survey,vandermaaten2009dimensionality,EspadotoSurvey}, thereby helping practitioners in choosing a suitable one for a given context.

Most projection techniques have been designed and evaluated only for \emph{static} data. Projecting dynamic (time-dependent) data is, however, equally important. Such data is found in most science and engineering areas, such as biology\,\cite{Teo2017}, medicine\,\cite{GRILLENZONI2019134}, and finance\,\cite{KRAPL2019101506}. The body of research in time series visualization is rich\,\cite{time0}, thereby underlining the importance of visualizing such data. Yet, there are only few examples of projecting time-dependent data\,\cite{Hu2010,Mao2007,Ward2011,Bernard2012,Nguyen2017,Jackle2016}. Even fewer works focus on designing projection techniques specifically for dynamic data\,\cite{Rauber2016,Fujiwara2019}. In particular, it is not clear how to measure \emph{and} trade-off two key aspects of such projections: \emph{visual quality} and \emph{stability}. While visual quality was studied well for static projections, stability, seen as the ability to create a set of projections that allows users to maintain a cohesive mental map through time, is recognized as essential for dynamic data visualization\,\cite{Archambault2011,Brehmer2019ACE}, but has not been formally defined nor quantified for dynamic projections.

We work towards filling this gap in assessing projection techniques for dynamic data with the following main contributions:
\begin{itemize}
  \item We propose novel variations of existing static projection \emph{techniques} for the context of visualizing time-dependent data;
  \item We propose a set of \emph{metrics} to quantify the stability of dynamic projections;
  \item We \emph{benchmark} the visual quality and stability of dynamic projections on a dataset collection to get insights on which methods favor which of the measured quality aspects.
\end{itemize}

Our work can help researchers in targeting the identified challenges of current dynamic projection techniques, therefore potentially leading to improved ones. Separately, practitioners can use our findings into the process of determining which dynamic projection technique is best suited to their given user context. Finally, our creation of an open benchmark for assessing dynamic projections (containing datasets, techniques, metrics, visualizations, and associated workflows) should benefit both user types by providing a basis via which such techniques can be transparently compared.

The structure of this paper is as follows. Section~\ref{sec:related} outlines related work and evaluation techniques for projections for static and dynamic data. Section~\ref{sec:experiment} details the proposed experiment we conducted to quantitatively assess the behavior of projection techniques for dynamic data, including techniques, datasets, and evaluated metrics. Section~\ref{sec:results} presents the obtained results. Section~\ref{sec:discussion} discusses the causes of the observed dynamic projection behavior. Finally, Section~\ref{sec:conclusion} concludes the paper. For replication purposes, all our datasets, code, workflow, and results are openly available\,\cite{repo}.

\section{Related work}
\label{sec:related}

\subsection{Preliminaries}
\label{sec:preliminaries}
We first introduce some notation. Let
$\mathbf{x} \in \mathbb{R}^n$
be an $n$-dimensional sample. A revision $\mathbf{R}^t = \{\mathbf{x}_i^t\}$, or timestep, of our data consists of a set of $N$ samples $\mathbf{x}_i^t$, $1 \leq i \leq N$ measured at the same time moment $t$. A dynamic dataset $\mathbf{D}$ is a list of $T$ revisions $\mathbf{D}=\left \{ \mathbf{R}^{t} \right \}, 1 \leq t \leq T$. For simplicity of exposition and implementation, but without loss of generality, we consider next that the sample count $N$ is constant over time. In this case, $\mathbf{D}$ can be represented as a set of $T$ $N$-by-$n$ matrices, one for each timestep.

A projection technique is a function $P: \mathbb{R}^{n} \rightarrow \mathbb{R}^{q}$, where $q \ll n$. For visualization purposes, $q \in \{2,3\}$. Since 2D projections are by far the most commonly used, we next only consider the case $q=2$. We denote the projection of observation $\mathbf{x}$ by $P(\mathbf{x})$. For each timestep $t$, let $P(\mathbf{R}^{t})$ be the 2D scatterplot of all points in
$\mathbf{R}^{t}$. Finally, let $P(\mathbf{D})$ be the set of $T$ scatterplots for all timesteps of dataset $\mathbf{D}$. These can be rendered as animations, small multiples, trail sets, or other visualization encodings.


Visualization of high dimensional data\,\cite{Liu2017} is a well studied topic populated with many techniques such as parallel coordinate plots\,\cite{Inselberg}, table lenses\,\cite{Rao2003}, scatterplot matrices\,\cite{Becker1996}, and dimensionality reduction (DR) methods\,\cite{nonato_survey,vandermaaten2009dimensionality,EspadotoSurvey}. From all these we next focus only on DR techniques, both for static and dynamic data, and evaluation methods for both of these technique classes.

\subsection{Techniques for \textit{static} dimensionality reduction}
The body of research that encompasses static DR is large and spans the fields of Information Visualization and Machine Learning. There are dozens of static techniques designed to optimize different objectives and to work well under different constraints. These can be classified and categorized using several taxonomies\,\cite{vandermaaten2009dimensionality} that guide users in choosing methods that meet their requirements. We do not further elaborate on such techniques, as several surveys extensively discuss static projections. Fodor \emph{et al.}\,\cite{fodor02_survey} present, to our knowledge, the first survey of DR techniques covering non-linear, vector quantization, and deep learning methods. Yin\,\cite{yin07_survey} surveys non-linear DR methods. Bunte \emph{et al.}\,\cite{bunte11} proposes a framework to quantitatively compare nine DR methods. Cunningham \emph{et al.}\,\cite{cunningham15_survey} presents a theoretical comparison of 15 linear DR techniques. A similar survey, extended to 30 DR techniques, both linear and non-linear, is provided by Sorzano \emph{et al.}\,\cite{sorzano14_survey}. Additional surveys look at DR methods in the larger context of high-dimensional data visualization, thus comparing and contrasting them with other visualization techniques\,\cite{buja96,hoffman02,engel12,kehrer13}. The most recent survey in this area\,\cite{nonato_survey} discusses technical aspects of DR methods, and also how such methods satisfy various user-level tasks.

\subsection{Evaluations of \emph{static} dimensionality reduction}
\label{sec:eval_static}
Taxonomies as the ones listed above, compare DR methods mainly from technical (algorithmic) and task-suitability aspects. An increasingly visible alternative approach is to compare techniques by measuring various quality \emph{metrics} on several techniques and datasets. A wealth of such quality metrics exist -- for recent overviews, see \cite{polzlbauer04_survey,Lee2009,lueks13,nonato_survey,EspadotoSurvey}. Different metrics gauge different desirable aspects of a projection, and usually, several metrics are jointly used to assess DR quality\,\cite{gisbrecht15}. Just as for DR techniques, metrics can be organized using different taxonomies. Following \cite{EspadotoSurvey}, these are as follows. \emph{Aggregate} metrics, such as trustworthiness, continuity, neighborhood hit, distance and class consistency\,\cite{sips09,tatu10}, cluster visual separation metrics\,\cite{albuquerque11,sedlmair13,sedlmair15}, and metrics that capture human perception based on machine learning\,\cite{sedlmair16} characterize an entire 2D scatterplot by a single scalar value. This is convenient when comparing (many) different scatterplots to choose a suitable one, such as in scagnostics applications. However, a scatterplot may exhibit different quality values in different areas, so a single aggregated value may not be suitable\,\cite{lamp,nonato_survey}. \emph{Point pair} metrics address this by measuring how point pairs $(P(\mathbf{x}), P(\mathbf{y}))$ in a projection relate to their corresponding sample pairs $(\mathbf{x}, \mathbf{y})$. These include Shepard diagrams\,\cite{lamp} and co-ranking matrices\,\cite{Lee2009}. Finally, \emph{local} metrics gauge separately every (small) neighborhood in a projection, thus providing the highest level of detail, and are typically visualized atop of the projection itself. These include the projection precision score\,\cite{schreck10}, stretching and compression\,\cite{aupetit07,lespinats11}, and false neighbors, missing neighbors, and average local errors\,\cite{Martins2014,Martins2015}.

Since all the above metrics aim to capture spatial aspects of the projection, we refer to them next as spatial quality metrics. Recent surveys have proposed extensive evaluations of spatial quality metrics on benchmarks containing a variety of datasets and DR methods\,\cite{EspadotoSurvey,vandermaaten2009dimensionality}. However, time-dependent datasets were not considered.

\subsection{Techniques for \emph{dynamic} dimensionality reduction}
The literature is much less rich regarding DR methods that \emph{explicitly} consider dynamic data. The dynamic t-SNE (dt-SNE) method of Rauber \emph{et al.}\,\cite{Rauber2016} extends the well-known t-SNE method\,\cite{tsne} by adding a stability factor $\lambda$ to the objective function. Such a factor jointly minimizes the Kullback-Leibler divergence proposed by t-SNE to preserve high-dimensional point neighborhoods and also restricts the amount of motion $\| P(\mathbf{x}^{t+1}) - P(\mathbf{x}^t) \|$ that points can have between consecutive timesteps. More recently, Fujiwara \emph{et al.}\cite{Fujiwara2019} proposed a PCA-based method to deal with streaming data. Note that this is a harder (and different) problem from the one we aim to study since one cannot anticipate changes occurring upstream in the data when optimizing for placement of points in 2D. As such, analyzing this (and similar) methods is out of our scope. Separately, several authors use DR methods to create static maps that describe multivariate time series. Hu \emph{et al.}\,\cite{Hu2010} use Self-Organizing Maps\,\cite{Kohonen1997} to create 2D trails that capture the dynamics of human motion data. Rauber \emph{et al.}\,\cite{rauber17} use similar trails, created by dt-SNE, to visualize the learning process of a neural network. Mao \emph{et al.}\,\cite{Mao2007} use PCA to project text feature evolution in text sequences. Ward and Guo\,\cite{Ward2011}, Bernard \emph{et al.}\,\cite{Bernard2012} and, more recently, Ali \emph{et al.}\,\cite{Ali2019} use similar approaches to find cyclic behavior, outliers, and trends in temporal data from medical, financial, and earth sciences domains.
In contrast to the previous methods, m-TSNE\,\cite{Nguyen2017} describes multivariate time series at a higher level of aggregation as single points instead of trails or polylines. Temporal MDS\,\cite{Jackle2016} projects $\mathbf{D}$ as a series of 1D projections, creating a map where the x-axis is time, and the y-axis shows the similarity of observations.

\subsection{Evaluation of \textit{dynamic} dimensionality reduction}
\label{sec:eval_dynamic}
Evaluating dynamic DR methods can be split into two aspects. First, just like for static DR methods, one is interested to see how well techniques capture the \emph{spatial} aspects of the underlying data. For this, one typically uses the same types of spatial quality metrics as for static projections (Sec.~\ref{sec:eval_static}). A separate important aspect for dynamic DR methods is \emph{stability}. Loosely put, stability describes how a dynamic DR technique encodes \emph{changes} in the data into \emph{changes} in the 2D metaphor used to visualize the data. Such metaphors can be grouped into spatial ones, where different timesteps map to different plots, such as in small multiples; and animation-based ones, where different timesteps are encoded into frames of a 2D animation.
Stability metrics were proposed and evaluated to assess the quality of other visualizations of dynamic data such as time-dependent treemaps\,\cite{sondag17,vernier18git,vernier18software}.

Stability is related to the capacity of a DR technique to deal with so-called out-of-core data. Simply put, this means the ability for a projection, created from a given dataset $\mathbf{D}$, to add extra points $\mathbf{X} \notin \mathbf{D}$ to the resulting 2D depiction $P(\mathbf{D})$, without distorting this depiction too much so that its understanding becomes hard. While recent works consider out-of-core and stability as key properties for DR projections\,\cite{nonato_survey,Boytsov2017,MateusEspadoto,Garcia-fernandez2013,Buja2008}, we are not aware of specific quality metrics that quantify these.

\section{Experimental setup}
\label{sec:experiment}
To evaluate how dynamic DR techniques perform, we follow a methodology similar to the one proposed in\,\cite{EspadotoSurvey} for evaluating static DR techniques, as follows.
We first select a set of dynamic DR \emph{techniques} to evaluate. Next, we select a collection of \emph{datasets} that cover various aspects, or \emph{traits}, that characterize high-dimensional dynamic data. Thirdly, we evaluate both spatial quality and stability \emph{metrics} on all combinations of techniques and datasets; in this step, we also propose novel metrics to gauge stability. We describe all these steps next. The analysis of the discovered correlations between techniques, dataset traits, and quality metrics obtained from our experiments is discussed afterwards in Sec.~\ref{sec:results}.

\subsection{Techniques}
\label{subsec:techniques}
We selected the dynamic DR techniques to evaluate based on the following considerations. First, we only consider techniques $P$, which, given a dataset consisting of several timeframes $\mathbf{R}^t$, produce corresponding 2D scatterplots $P(\mathbf{R}^t)$. We argue that this is the most generic definition of a dynamic projection -- from such scatterplots, other types of visualizations can be constructed next as desired (animation, small multiples, trails). This is analogous to expecting a generic static projection technique to deliver a 2D scatterplot. Hence, techniques that deliver different output types, such as m-TSNE\,\cite{Nguyen2017} and temporal MDS\,\cite{Jackle2016}, are excluded from our evaluation. Secondly, we only consider techniques that (1) are generic with respect to the input data (size, dimensionality, provenance) they can handle; (2) well-known and often used in practice, so their evaluation arguably serves a sizeable user group; and (3) easy to set up, control, and have publicly available implementations, for reproducibility. We next describe the selected techniques.

\noindent\textbf{t-SNE and variants:} Probably the simplest way to project dynamic data is to compute a single, global, projection $P(\mathbf{D})$ for the entire dataset $\mathbf{D}$ and next visualize the timeframes by using the desired method, be it animation, trails, or small multiples. We next call this the \emph{global} (G) approach. While this arguably favors stability (since $P$ sees all data $\mathbf{D}$ at once), it likely yields limited spatial quality, since $P$ has the challenging task of placing well \emph{all} points from \emph{all} revisions in $\mathbf{D}$. An equally simple approach is to compute independent projections $P(\mathbf{R}^t)$ for each revision $\textbf{R}^t$. We call this next the \emph{per-timeframe} (TF) approach. This arguably favors spatial quality, since $P$ must only optimize positions for each revision $\mathbf{R}^t$ separately, rather than the entire $\mathbf{D}$. However, this approach can yield poor stability, since timeframes are projected without knowledge of each other. Both the global and timeframe approaches were suggested, but not quantitatively evaluated, in the dt-SNE paper\,\cite{Rauber2016}. Given this, and also the fact that t-SNE is a very well-known static technique, we next consider G-t-SNE, TF-t-SNE, and dt-SNE in our evaluation.

\noindent\textbf{UMAP}: This recent DR technique\,\cite{umap} has a mathematical foundation on Riemannian geometry and algebraic topology. According to recent studies\,\cite{EspadotoSurvey,Becht2019}, UMAP offers high-quality projections with lower computational cost and better global structure preservation than t-SNE, being thus an interesting competitor in the DR arena.  We consider in our evaluation both the global (G-UMAP) and per-timeframe (TF-UMAP) variants of this technique.

\noindent\textbf{PCA:} Following \cite{Fujiwara2019,Mao2007,Ward2011}, we also consider Principal Component Analysis\,\cite{pca}, implementing the global and timeframe strategies. In detail, PCA performs a linear mapping of the data $\mathbf{D}$ to, in our case, 2D by maximizing the data variance in the 2D representation. The global strategy implies computing PCA once for the entire $\mathbf{D}$. In contrast, timeframe PCA means computing PCA separately for each revision $\mathbf{R}^t$. Given the widespread use of PCA in many fields of science, and also its out-of-core ability (which, as outlined in Sec.~\ref{sec:eval_dynamic}, is related to stability), we consider both G-PCA and TF-PCA next in our evaluation.

\noindent\textbf{Autoencoders:} Often used in dimensionality reduction and representation learning, autoencoders\,\cite{aes,Ballard1987} are hourglass-shaped neural networks. They are composed of an encoder that takes the original data $\mathbf{D}$ and compresses it into a compact (latent) representation $P(\mathbf{D})$ of lower dimensionality (two in our case), and a decoder, which takes $P(\mathbf{D})$ and aims to reconstruct a good approximation of the original data $\mathbf{D}$. While autoencoders have been often used to create static projections of high-dimensional data, they have not, to our knowledge, been quantitatively evaluated for their ability to handle dynamic data. We evaluated four types of autoencoders, as follows. \emph{Dense autoencoders} (AE) are comprised of only fully-connected (dense) layers and are the standard variant. \emph{Convolutional autoencoders} (CAE)\,\cite{Masci2011} have both fully-connected and convolutional layers. The convolutional layers apply a non-linear transformation to the data that takes into account the spatial correlation between attributes, for instance, the proximity of pixels in an image. \emph{Variational autoencoders} may have both fully-connected layers (VAE)\,\cite{Kingma2013} and convolutional layers (CVAE). The main difference between dense and variational autoencoders is the addition of stochastic behavior in the intermediate layer of the latter. The encoder produces two vectors -- an intermediate representation (IR) and an uncertainty degree $\sigma$ for each IR value. The decoder tries to reconstruct the input through a sample from the latent space distribution with mean IR and variance $\sigma$, thus forcing the network to learn similar representation for similar inputs.
Convolutional based architectures are not generic regarding input and a meaningful spatial relationship between attributes is expected (such as found on image data). We, therefore, restrain the analysis on this document to AE and VAE. The results of CAE and CVAE runs for the image based datasets (fashion and quickdraw) can be found on the online material\,\cite{repo}.

\noindent\textbf{Implementation:} We implemented the chosen dynamic DR techniques (G-t-SNE, TF-t-SNE, dt-SNE, G-UMAP, TF-UMAP, G-PCA, TF-PCA, AE, CAE, VAE, CVAE) as follows. For t-SNE and PCA, we used scikit-learn\,\cite{scikit-learn} with default parameters. For dt-SNE and UMAP, we used the implementation provided online by the authors\,\cite{Rauber2016,umap}. Finally, we implemented the four autoencoder models using Keras\,\cite{chollet2015keras}, with different numbers of layers, nodes per layer, optimizers, and training routines. Tab.~\ref{tab:hyperparams} shows the values, for each autoencoder and dataset, that delivered the best results, and which we used next. The code, notebooks, and instructions to recreate our results are available online\,\cite{repo}.

\begin{table}[tb]
\centering
\fontfamily{lmss}\selectfont
\scriptsize
\caption{Hyperparameters of the autoencoder-based DR methods}
\label{tab:hyperparams}
\begin{tabular}{llcll}
\hline
dataset    & technique & \# hidden layers & \begin{tabular}[c]{@{}l@{}}\# nodes/layer\end{tabular} & \# epochs \\ \hline
cartolastd & AE  & 2                & 10, 10                      & 50        \\
cartolastd & VAE  & 2                & 10, 10                      & 100       \\
cifar10cnn & AE  & 2                & 10, 10                      & 20        \\
cifar10cnn & VAE  & 2                & 100, 10                     & 20        \\
esc50      & AE  & 2                & 10, 10                      & 40        \\
esc50      & VAE  & 2                & 100, 10                     & 20        \\
fashion    & AE  & 3                & 500, 500, 2000              & 40        \\
fashion    & VAE  & 3                & 2048, 1024, 512             & 20        \\
gaussians  & AE  & 2                & 10, 10                      & 20        \\
gaussians  & VAE  & 2                & 100, 10                     & 20        \\
nnset      & AE  & 2                & 10, 10                      & 20        \\
nnset      & VAE  & 2                & 100, 10                     & 20        \\
qtables    & AE  & 2                & 10, 10                      & 20        \\
qtables    & VAE  & 2                & 100, 10                     & 20        \\
quickdraw  & AE  & 3                & 500, 500, 2000              & 40        \\
quickdraw  & VAE  & 3                & 2048, 1024, 512             & 20        \\
sorts      & AE  & 2                & 10, 10                      & 20        \\
sorts      & VAE  & 2                & 100, 10                     & 20        \\
walk       & AE  & 2                & 10, 10                      & 20        \\
walk       & VAE  & 2                & 100, 10                     & 20        \\ \hline
\end{tabular}
\vspace{-0.15cm}
\end{table}

\subsection{Datasets}
\label{subsec:datasets}
There is, to our knowledge, no standardized benchmark for evaluating DR techniques. Espadoto \emph{et al.}\,\cite{EspadotoSurvey} took a first step towards providing such a benchmark containing 19 datasets. However, all these are time-independent, thus not suitable for us. We followed here a similar approach, \emph{i.e.} collecting a set of 10 high-dimensional and dynamic datasets that exhibit significant variations in terms of provenance, number of samples $N$, number of timesteps $T$, dimensionality $n$, intrinsic dimensionality $\rho_n$ (percentage of $n$ dimensions that explain 95\% of the data variance), and sparsity ratio $\sigma_n$ (percentage of zeros in the data). All datasets are labeled into 3 to 10 classes. We only use labels for visualization and quality assessment and not the projection itself. Table \ref{tab:datasets} shows the characteristics, or traits, for these datasets. Further details on them are listed below.

\begin{itemize}
  \item \textbf{cartolastd:} Player statistics for the second turn of the 2017 Brazilian football championship. Data was extracted from an open-source project\,\cite{cartola} that scrapes the Cartola FC football platform. Each timestep corresponds to a tournament round. Variables relate to per-match performance of a given player (number of goals, assistances, fouls, defenses, etc.). Players are labeled by their playing position (goalkeeper, right or left-back, defender, midfielder, forward).
  \item \textbf{cifar10cnn:} Last hidden layer activations after each training epoch for a convolutional network trained to classify the CIFAR10\,\cite{cifar10} dataset.
  \item \textbf{esc50:} Sound samples of 8 classes (brushing teeth, chainsaw, crying baby, engine, laughing, rain, siren, wind) compressed to 128 frequencies and smoothed over time. Extracted from Piczak's ESC50 dataset\,\cite{esc50}.
  \item \textbf{fashion:} 100 images from each of the 10 classes (T-shirt/top, trouser, pullover, dress, coat, sandal, shirt, sneaker, bag, ankle boot) of the FashionMNIST\,\cite{Xiao2017} dataset with decreasing amounts of noise over time.
  \item \textbf{gaussians:} Synthetic dataset used to evaluate dt-SNE\,\cite{Rauber2016}. Isotropic gaussian blobs in $nD$ with diminishing spread over time.
  \item \textbf{nnset:} Internal states (weights and biases) of several neural networks during training with the MNIST dataset\,\cite{lecun-mnist}. The networks have the same architecture but use different optimizers, batch sizes, and training-set sizes.
  \item \textbf{quickdraw:} Drawing sequences for 600 objects of 6 different classes drawn by random people. Extracted from the ``Quick, Draw!'' Google AI experiment\,\cite{quickdraw}.
  \item \textbf{sorts:} This dataset was designed to compare the behavior of eight sorting algorithms. The algorithms sort different arrays of 100 random values in $[0,1]$. As they do so, we take snapshots of the intermediate states, until sorting is over. Each observed point is an (algorithm, array) run, and its feature vector is the partially sorted array at a given time.
  \item \textbf{walk:} Synthetic dataset with similar structure to \emph{gaussians}. It contains 3 high-dimensional clusters oscillate (approach, intermingle and cross, and then drift apart) in $\mathbf{R}^{100}$ over time. We designed this dataset to see how well the studied DR techniques can capture the approaching, mingling, and drifting-away dynamics mentioned above.
\end{itemize}


Covering all variations of high-dimensional datasets with a benchmark is already daunting for static data\,\cite{EspadotoSurvey}, thus even more for dynamic data, as there are many types of dynamic patterns possible. Hence, we cannot claim that our benchmark is \emph{exhaustive} in terms of the space it samples. However, we believe that the included datasets exhibit a rich variety of different traits (Tab.~\ref{tab:datasets}). Also, no two datasets are redundant, \emph{i.e.}, have all traits similar. Given that, to date, no other benchmark exists for this task, we believe ours is a good start in supporting the intended evaluation.

\begin{table}
\centering
\caption{Datasets and their traits used in the evaluation.}
\label{tab:datasets}
\vspace{-0.15cm}
\includegraphics[width=.99\linewidth]{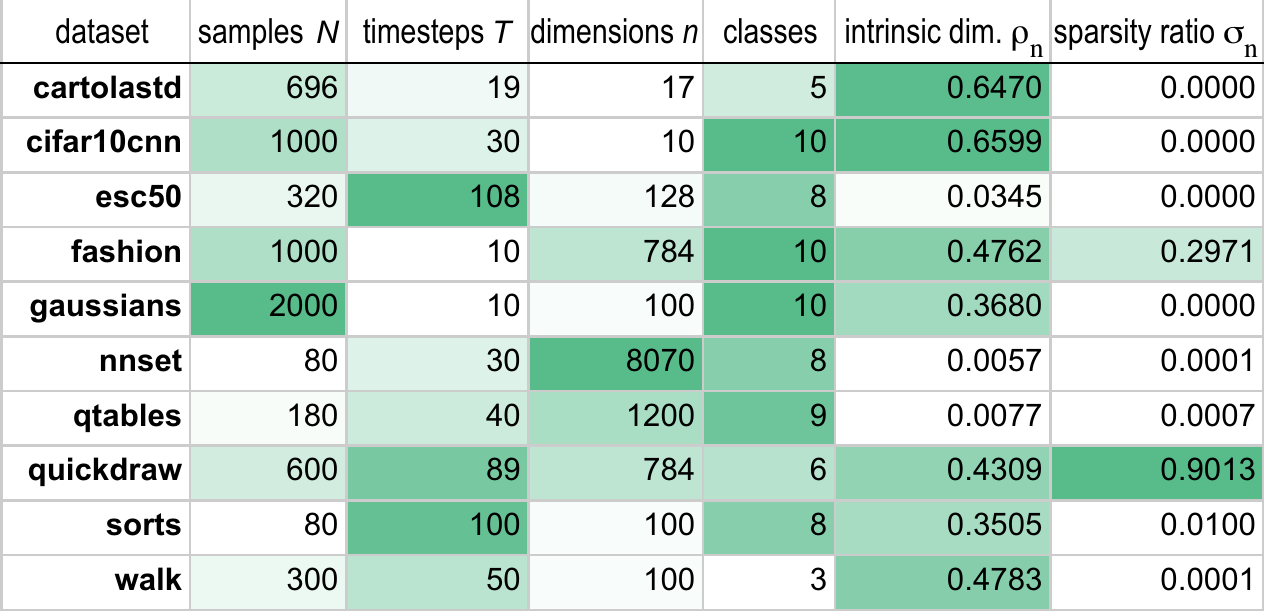}
\vspace{-0.15cm}
\end{table}

\vspace{-0.15cm}
\subsection{Metrics}
\label{subsec:metrics}
We measure the quality of all projection techniques (Sec.~\ref{subsec:techniques}) on all datasets (Sec.~\ref{subsec:datasets}) using both spatial quality and stability metrics, similarly to other evaluations of multivariate dynamic data visualizations such as treemaps\,\cite{sondag17,vernier18git,vernier18software}. In our evaluation, we use the same metrics as the survey\,\cite{EspadotoSurvey} (and a few extra ones) over all revisions $\mathbf{R}^t$, as follows.

\subsubsection{Spatial metrics}
\label{sec:spatial}
\noindent
\textbf{Neighborhood preservation ($S_{NP}$):} With values in $[0,1]$, with 1 being the best, this is the percentage of the k-nearest neighbors of $\mathbf{x} \in \mathbf{D}$ that project in the k-nearest neighborhood of $P(\mathbf{x})$.

\noindent
\textbf{Neighborhood hit ($S_{NH}$):} With values in $[0,1]$, with 1 being the best, this is the fraction of the k-nearest neighbors of a projected point $P(\mathbf{x})$ that have the same class label as $P(\mathbf{x})$.
Since we know that our datasets exhibit reasonably well-separated classes in $\mathbb{R}^n$, a proper DR technique (from the perspective of class separation tasks) should yield a high neighborhood hit.

\noindent
\textbf{Trustworthiness ($S_{Trust}$):} With values in $[0,1]$, with 1 being the best, this measures how well the $k$ nearest neighbors $NN^k(P(\mathbf{x}))$ of a projected point $P(\mathbf{x})$ match the $k$ nearest neighbors $NN^k(\mathbf{x})$ of a data point $\mathbf{x}$. Simply put, trustworthiness measures how few missing neighbors\,\cite{Martins2014} a projected point has. Formally, if $U^k(\mathbf{x})$ is the set of points that project in $NN^k(P(\mathbf{x}))$ but are not in $NN^k(\mathbf{x})$,
and $r(\mathbf{x},\mathbf{y})$ is the rank of $\mathbf{y}$ in the ordered set of nearest neighbors $NN^k(P(\mathbf{x}))$, trustworthiness is then defined as \linebreak
$1-\frac{2}{N k(2 N-3 k-1)} \sum_{x=1}^{N} \sum_{y \in U^k(\mathbf{x})}(r(x, y)-k)$.

\noindent
\textbf{Continuity ($S_{Cont}$):} With values in $[0,1]$, with 1 being the best, this measures how many missing neighbors\,\cite{Martins2014} a projected point has. Following the above notations, let $V^k(\mathbf{x})$
be the points that are in $NN^k(\mathbf{x})$ but do not project in $NN^k(P(\mathbf{x}))$. Let also $\hat{r}(\mathbf{x}, \mathbf{y})$ be the rank of $\mathbf{y}$ in the ordered set of neighbors $NN^k(\mathbf{x})$. Continuity is then defined as \linebreak
$1-\frac{2}{N k(2 N-3 k-1)} \sum_{x=1}^{N} \sum_{y \in V^k(\mathbf{x})}(\hat{r}(x, y)-k)$.

In contrast to\,\cite{EspadotoSurvey}, we compute neighborhood preservation, trustworthiness, and continuity for multiple (20) neighborhood sizes equally spread between $k=1\%$ and $k=20\%$ of the point count $N$. Similarly, for the neighborhood hit, we use 20 values for $k$, ranging from 0.25\% to 5\%. This allows us next to study the spatial quality of projections at different scales\,\cite{Martins2015}.

\noindent
\textbf{Normalized stress ($S_{Stress}$):} With values in $\mathbb{R}^{+}$, lower meaning better distance preservation, stress measures the pairwise difference of distances of points in $nD$ and $qD$. We define $S_{Stress}$ as  $\sum_{i j}\left(d_{i j}^t - \overline{d_{i j}^t}\right)^{2} / \sum_{i j} (d_{i j}^t)^{2}$,
where $d_{i j}^t$ and $\overline{d_{i j}^t}$ are the Euclidean distances between data points $\mathbf{x}_i^t$ and $\mathbf{x}_j^t$, and between their projections $P(\mathbf{x}_i^t)$ and $P(\mathbf{x}_j^t)$, respectively, for $1 \leq t \leq T$, for every point pair $(i,j)$. To ease analysis, we scale distances using standardization.

\noindent
\textbf{Shepard diagram metrics:} The Shepard diagram is a scatterplot of $d_{ij}$ by $\overline{d_{ij}}$, for every pair $(i,j)$ in $\mathbf{D}$ (see Fig. \ref{fig:trails_cartolastd}b). It visually tells how different ranges of distances between points are affected by a projection. Plots close to a diagonal indicates good distance preservation.
Deviations from this highlight patterns such as poor preservation of long/short distances, creation of false neighborhoods, or stretching and compression of the manifold on which the data is defined\,\cite{lamp}. We summarize and quantify Shepard diagrams by measuring the relationship between the two distances. Following \cite{EspadotoSurvey}, we use Pearson correlation to measure the linearity of the relationship, and we add Spearman and Kendall correlation to measure the monotonicity of the relationship. The three resulting correlation metrics $S_{Pearson}, S_{Spearman}, S_{Kendall}$ range from -1 to 1, where 1 means perfect positive correlation.

\subsubsection{Temporal stability metrics}
As previously stated, there are no metrics in the literature specially designed to measure the temporal stability of DR methods. We next propose two such metrics, as follows. The two variables whose relationship we want to measure are the \emph{change of the attributes} of a sample $\mathbf{x}$ from time $t$ to $t+1$, measured as the $n$D Euclidean distance
$ \delta^t = \|\mathbf{x}^t - \mathbf{x}^{t+1}\|$, and \emph{movement of the projection point $P(\mathbf{x})$} from time $t$ to $t+1$, measured as the 2D Euclidean distance
$ \overline{\delta^t} = \|P(\mathbf{x}^t) - P(\mathbf{x}^{t+1})\|$. Ideally, for a temporally stable $P$, we want $\overline{\delta^t}$ to be proportional to $\delta^t$. However, this may be a too hard constraint for $P$ to satisfy, just as perfect $n$D to 2D distance preservation is hard to achieve for static projections. A more relaxed requirement for a temporally stable $P$ is to have
 $\overline{\delta^t}$ a monotonic increasing function of $\delta^t$. Indeed, if this constraint were not obeyed by $P$, then if an observation $\mathbf{x}^t$ changes only slightly over time, its projection  $P(\mathbf{x}^t)$ could move a lot. That is, if  $\delta^t \ll \overline{\delta^t}$, the projection $P$ is unstable, and would convey the user the wrong impression that data is changing a lot. Conversely, if $\mathbf{x}^t$ strongly changes over time, but $P(\mathbf{x}^t)$ remains roughly static, \emph{i.e.} if $\delta_i^t \gg \overline{\delta_i^t}$, then the user gets the wrong impression that the data is not changing. Hence, for a temporally stable $P$, the two changes $\overline{\delta^t}$ and ${\delta^t}$ should be positively correlated.

To measure the relationship of ${\delta^t}$ and $\overline{\delta^t}$, we adapt the static spatial quality metrics introduced in Sec.~\ref{sec:spatial} as follows:

\noindent
\textbf{Normalized temporal stress ($T_{Stress}$):} We define temporal stress as $\sum_{i\, t}{(\delta_{i}^{t}-\overline{\delta}_{i}^{t})^{2}} / { (\delta_{i}^t)^{2}}$, where the subscript $i$ indicates sample point $\mathbf{x}_i$. As for the spatial normalized stress, we normalize distances using standardization. Low stress values indicate that the 2D changes $\overline{\delta^t}$ reflect closely their $n$D counterparts ${\delta^t}$, which is desirable.

\noindent
\textbf{Temporal Shepard diagram metrics:} Akin to the spatial metrics defined on Shepard diagrams, we measure the Pearson, Spearman, and Kendall correlations $T_{Pearson}, T_{Spearman}, T_{Kendall}$
between $\delta$ and $\overline{\delta}$ for every observation and consecutive timesteps. High correlation values indicate that the 2D changes $\overline{\delta^t}$ are strongly correlated with their $n$D counterparts ${\delta^t}$, which is desirable.

\section{Evaluation and Results}
\label{sec:results}
We evaluate the 12 quality metrics introduced in Sec.~\ref{subsec:metrics} on all (dataset, method) pairs formed by the selected 9 DR methods and 10 datasets, and analyze next the results. We do this by proposing several metric visualizations, from highly aggregated (to help forming first insights) to detailed (to examine more subtle points). For a direct impression, see also the videos showing the actual dynamic projections in action, available online at\,\cite{repo}.

\subsection{Aggregated results}
\label{sec:aggregate}

\begin{figure}[tb!]\centering
  \includegraphics[width=1.0\linewidth]{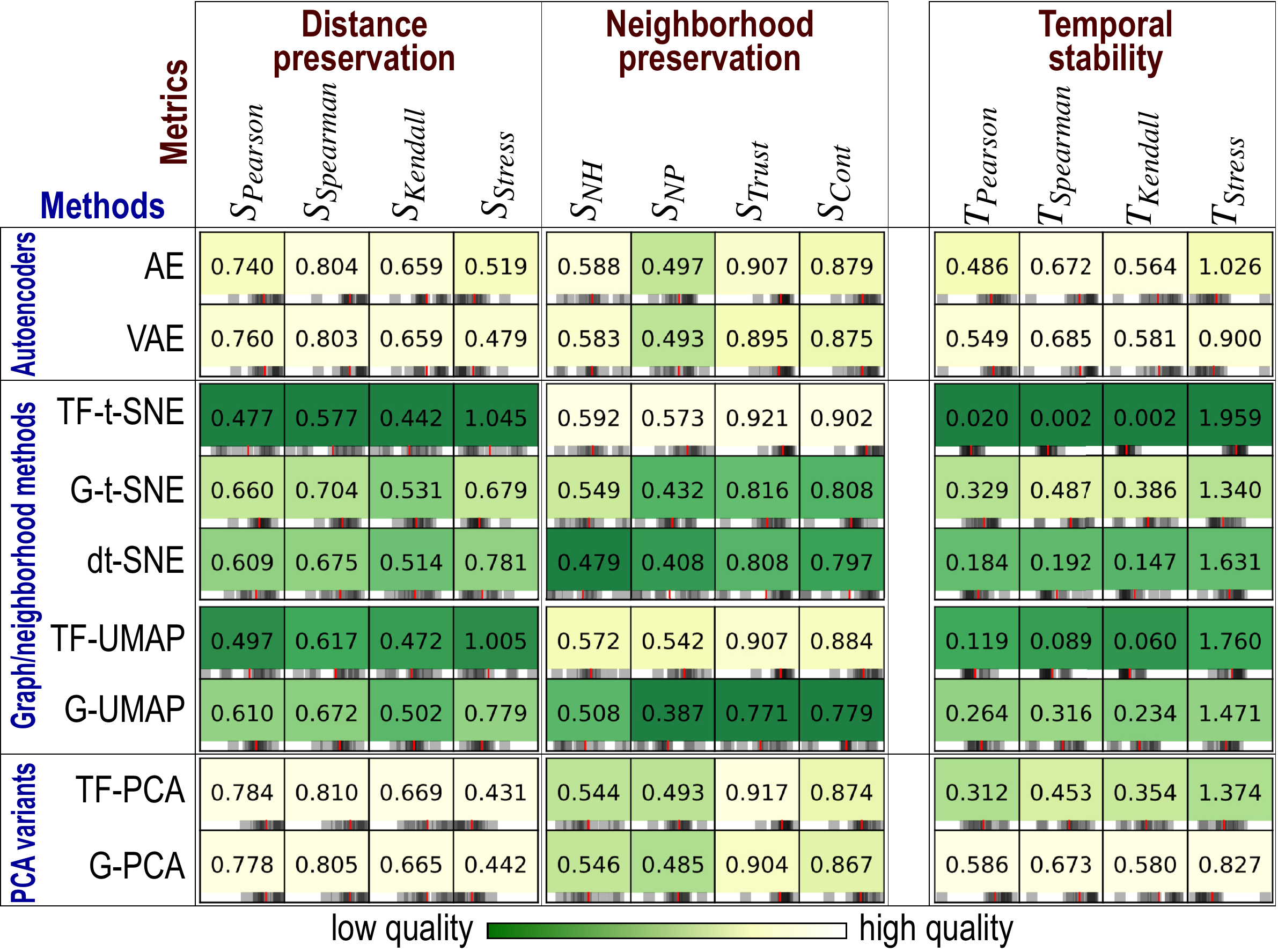}
  \caption{Aggregated metric results over all datasets.}
  \label{fig:aggregated}
\end{figure}

Figure~\ref{fig:aggregated} shows average metric values computed over all datasets and techniques.
Light colors represent high metric values (preferred). The colormap in Fig.~\ref{fig:aggregated} was normalized independently by the min and max of each column (metric), and it was inverted for the stress-based metrics, as low values mean preferred results for these. At the bottom of each cell, a 1D scatterplot with density mapped to luminance shows the distribution of the values of the (metric, method) pair corresponding to that cell over all datasets. The red line shows the distribution mean. The table in Fig.~\ref{fig:aggregated} is divided into three blocks: The two left blocks show spatial metrics for distance and neighborhood preservation, respectively. The right block shows stability metrics.

\begin{figure*}[!tb]\centering
  \vspace{-0.2cm}
  \includegraphics[width=.92\linewidth]{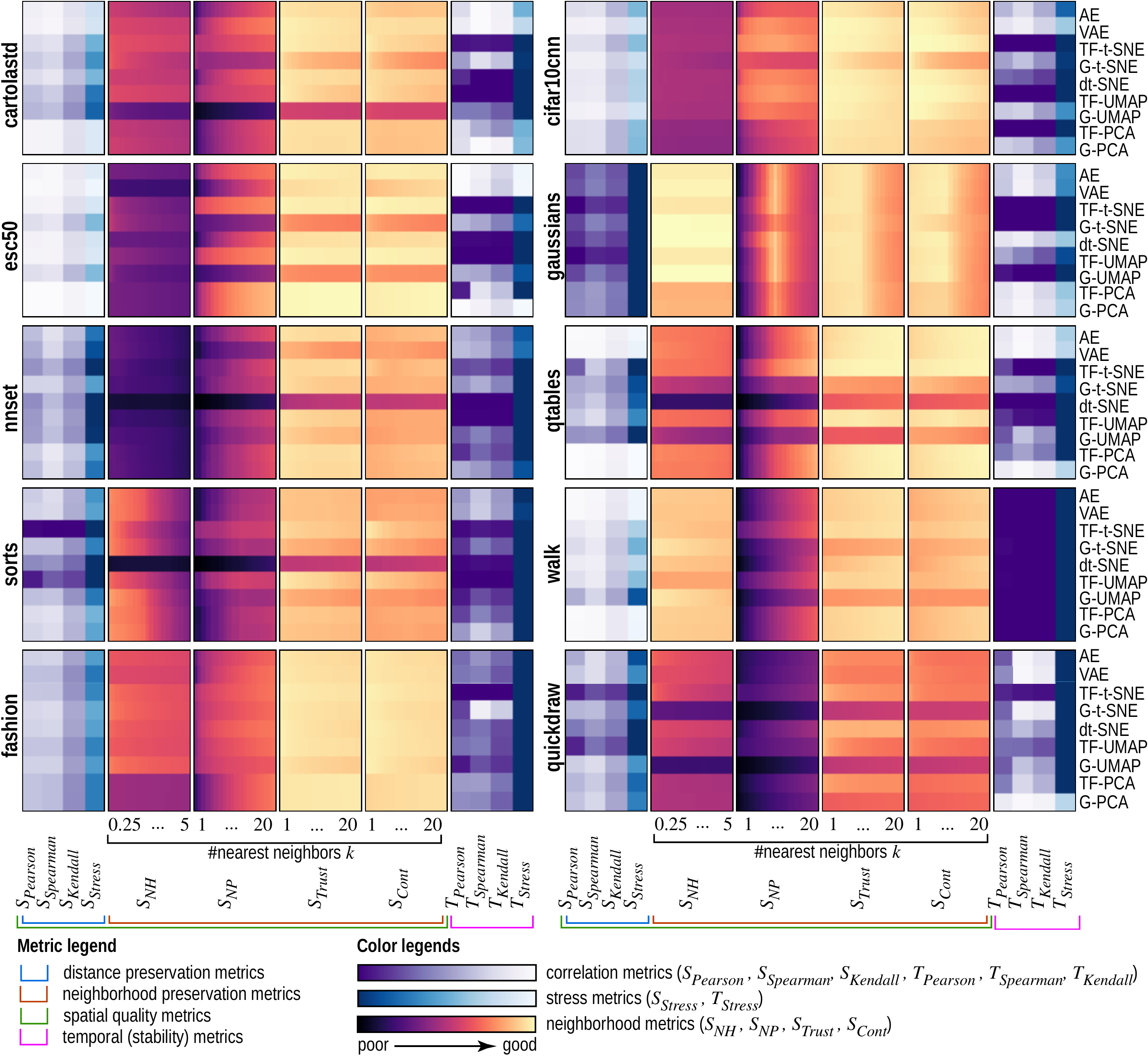}
  \caption{Twelve spatial quality and temporal stability metrics evaluated for 9 DR  methods run on ten datasets.}
  \vspace{-0.2cm}
  \label{fig:all_datasets}
\end{figure*}

Figure~\ref{fig:aggregated} helps us to find methods that strike a balance between spatial quality and stability. In this sense, (variational) autoencoders and G-PCA score, overall, the best. The other methods are good in one aspect but not the other: Timeframe t-SNE has high neighborhood metric values but poor distance preservation and the poorest stability from all assessed methods. Timeframe PCA has high distance preservation but relatively low stability. dt-SNE appears to be as good spatially as G-t-SNE, but slightly less stable. This is an interesting finding since dt-SNE was explicitly designed (but not quantitatively assessed) to aid stability.

\subsection{Dataset-wise results}
Figure~\ref{fig:aggregated} is simple to read but heavily aggregated, so it does not show how the quality of specific methods depends on specific \emph{datasets}. To see this, Fig.~\ref{fig:all_datasets} shows all metric results for all datasets without aggregation. As in Fig.~\ref{fig:aggregated}, light colors mean good results. Columns are now not normalized. Column groups (a-f) represent spatial metrics, and columns (g-h) represent stability metrics. We use different quantitative colormaps to indicate different types of measured data. By examining Fig.~\ref{fig:all_datasets}, we obtain the following insights:

\noindent\textbf{Unstable methods:}  TF-t-SNE is always unstable regardless of the dataset. This refines the instability finding over TF-t-SNE (Sec.~\ref{sec:aggregate}) by showing that this occurs irrespective of the dataset. Also, it confirms the same observation in \cite{Rauber2016}, which, however, was not quantitatively confirmed there. The reason for this instability is the stochastic nature of t-SNE, which strongly manifests itself if we run the method from scratch on every new revision (timeframe). We could attribute the instability of TF-UMAP to the same reason.

\noindent\textbf{Poor spatial quality:} G-t-SNE and G-UMAP score poorly on distance and neighborhood preservation on most datasets. This is the aforementioned difficulty (Sec.~\ref{subsec:techniques}) of constructing a \emph{single} projection covering many samples in many timeframes. This is much harder than constructing a projection that preserves only neighborhoods formed by points in a \emph{single} timeframe. We see here again the trade-off between spatial quality and stability.

\noindent\textbf{Neighborhood preservation:}
Here we see dataset-specific behavior: For \emph{gaussians}, $S_{NP}$, $S_{Trust}$, and $S_{Cont}$ peak at a neighborhood size of roughly 10\% of the dataset size. This makes sense since this is the size of the clusters present in this dataset -- when $k$ exceeds this value, the metrics will start considering points in other clusters, thus decrease. More interestingly, we see some outliers (dark bands in the heat-colormapped plots). These are techniques that score poorly for any $k$ value. Among these, we find G-t-SNE, dt-SNE, and G-UMAP. At the other extreme, TF-t-SNE and TF-UMAP score the best results at neighborhood preservation, followed by AE, VAE, G-PCA, and TF-PCA.

\noindent\textbf{Dynamic t-SNE:} In contrast to the good results qualitatively observed on the single \emph{gaussians} dataset showed in \cite{Rauber2016}, dt-SNE performs less well in both spatial quality and stability for several other of the considered datasets, being quality-wise somewhere between TF-t-SNE and G-t-SNE for all considered metrics.

%

\noindent\textbf{Dataset difficulty:} Some datasets are considerably harder to project with good quality than others, no matter which technique we use. For example, \emph{walk} has poor stability for all techniques. In contrast, \emph{gaussians} has good stability for all techniques (except the t-SNE and UMAP variants) and good neighborhood preservation for all techniques. To study how dataset characteristics influence quality, we compute the correlation of the distance-preservation, neighborhood, and temporal stability metrics (measured over all techniques) with the six traits that we used to characterize our datasets (Tab.~\ref{tab:datasets}). Table~\ref{tab:corr_table} shows the results. A few things stand out: As the number of samples $N$ increases, the difficulty to preserve distances also increases, but neighborhoods are preserved better. Conversely, as sparsity $\sigma_n$ increases, it becomes harder to preserve neighborhoods. Separately, we do not find any strong (positive or negative) correlation of temporal stability with any of the traits. Overall, this suggests that the traits are useful in predicting \emph{spatial} quality of projections. However, we need additional traits that capture the data dynamics to reason about the projections' temporal stability.

\begin{table}[!tb]
\centering
\caption{Correlation between metric types and dataset traits.}
\label{tab:corr_table}
\includegraphics[width=1.01\linewidth]{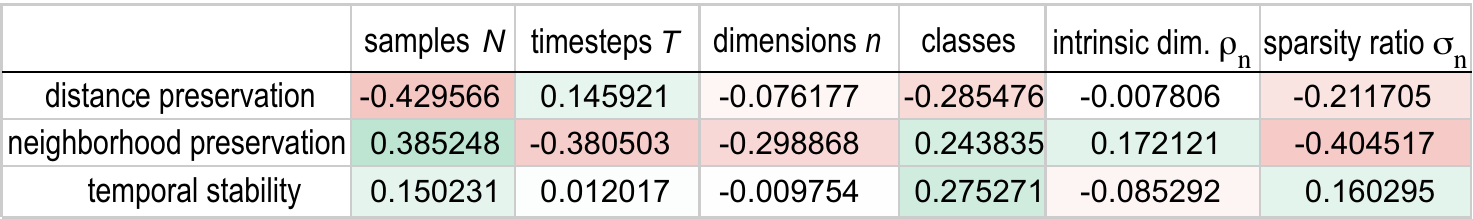}
\end{table}

\begin{figure*}[!tb]\centering
\hspace*{-0.02\linewidth}
  \includegraphics[width=1.0\linewidth]{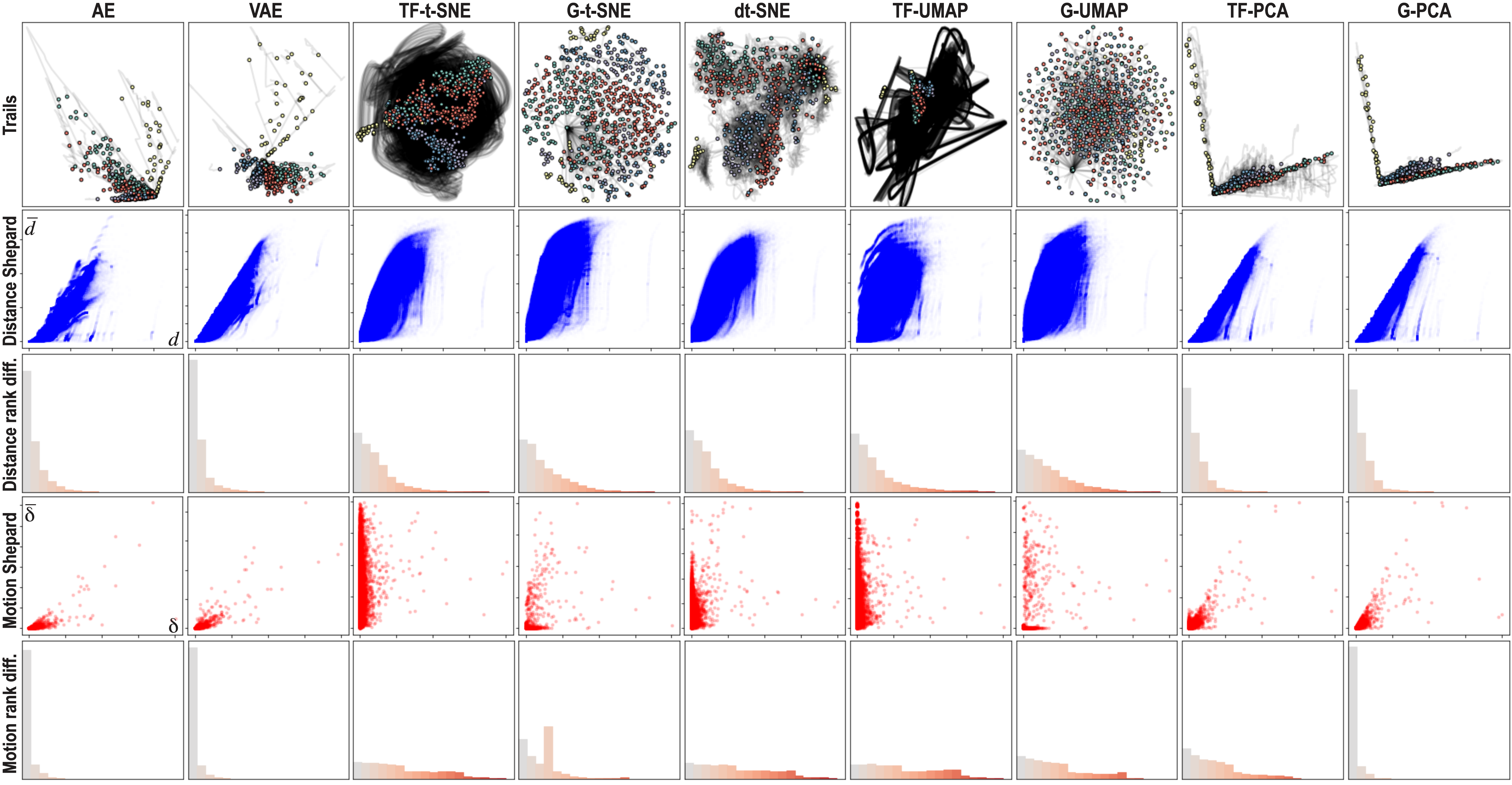}
  \caption{Detailed analysis of distances and movements produced by all DR techniques on the \emph{cartolastd} dataset.}
  \vspace{-0.15cm}
  \label{fig:trails_cartolastd}
\end{figure*}

\begin{figure*}[tb!]\centering
  \includegraphics[width=0.85\linewidth]{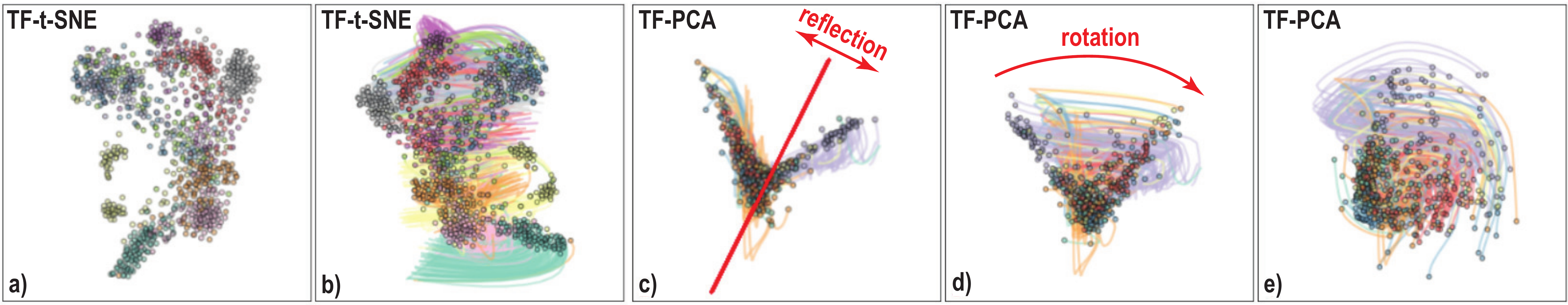}
  \caption{Examples of instability in TF-t-SNE (a,b) and TF-PCA (c,d,e).}
  \vspace{-0.15cm}
  \label{fig:instability}
\end{figure*}

\subsection{Fine-grained analysis}
While Fig.~\ref{fig:all_datasets} shows all computed metrics for each (dataset, method) combination, metric values are still aggregated to a single scalar per combination. This does not show how metrics vary over the \emph{extent} of a projection and/or over \emph{time}. There are more patterns in dynamic projections than we can capture by a set of metrics, no matter how good these are. To get such insights, we next present a fine-grained analysis that aggregates the metrics even less (see Figure~\ref{fig:trails_cartolastd}) for a single dataset (\emph{cartolastd}, chosen as it is alphabetically the first in our benchmark).
Similar visualizations for all other datasets in the benchmark are available online\,\cite{repo}. We next analyze these methods for this dataset from several perspectives, as follows.

\noindent\textbf{Stability visual assessment:} Figure~\ref{fig:trails_cartolastd}a shows the actual dynamic projections with point trails $(P(\mathbf{x}_i^1),\ldots, P(\mathbf{x}_i^T))$, one per player $i$. Colors map the players' labels. This visualization already says a lot about the behavior and similarities of the studied DR methods (see also the submitted videos). The instability of TF-t-SNE and TF-UMAP becomes apparent, as their trails cover a very large area in the projection space. However, these methods achieve a quite good separation of same-label clusters. In contrast, dt-SNE shows trails that depict much local movement. Both PCA variants show relatively little movement, with points oscillating along two main axes, which are the main eigenvectors computed by the methods. At the other extreme, AE, VAE, and G-t-SNE show the least motion. However, this does not imply by itself a high quality: G-t-SNE, for instance, achieves indeed a better visual spreading of samples in the available projection space, but it has very poor neighborhood preservation (see G-t-SNE results in Fig.~\ref{fig:all_datasets}) and, as already discussed above, it also has very poor stability.

\noindent\textbf{Distance preservation:} Figure~\ref{fig:trails_cartolastd}b shows the Shepard diagram of distances, which is a scatterplot of $d_{ij}$ by $\overline{d_{ij}}$, for every pair $(i,j)$ in $\mathbf{D}$, that helps us understand the distance preservation aspect of each technique. We see that the AE and PCA variants have overall better distance preservation (plots closer to the diagonal) than the t-SNE/UMAP variants. Also, we see that AE and PCA typically \emph{compress} $n$D distances to 2D (points mainly under the main diagonal), whereas the t-SNE/UMAP variants both compress and stretch these (points are located both under and above this diagonal).

Inspired by the Spearman and Kendall correlations, we consider next the agreement of \emph{ranks} instead of aggregating it to a single value. Figure~\ref{fig:trails_cartolastd}c shows this, for distance preservation, by a histogram of the \emph{absolute} rank differences of $n$D and 2D distances between point pairs. In a projection with $S_{Spearman} = S_{Kendall} = 1$, such differences would be minimized, \emph{i.e.}, the $k^{th}$ largest 2D distance $\overline{d_{ij}}$ should correspond to the $k^{th}$ largest $n$D distance $d_{ij}$ for every point pair $(i,j)$. In this case, all rank differences are zero, which would yield a histogram showing a single high bar at zero (left of the histogram). Significant rank differences spread the histogram to the right, showing poor monotonicity between the two variable ranks. From these plots, we see, again, that AE and VAE score the best, followed by G-PCA, TF-PCA, and then the t-SNE and UMAP variants.

\noindent\textbf{Stability metrics:} Figure~\ref{fig:trails_cartolastd}d shows Shepard diagrams for the point movements, \emph{i.e.}, scatterplots of $\delta$ by $\overline{\delta}$ for every sample compared to itself in the next timestep, for all timesteps. Note that, in these scatterplots, every point is a \emph{sample}, whereas in the classical Shepard diagrams (Fig.~\ref{fig:trails_cartolastd}b), every point is a \emph{pair} of samples. Ideally, we want $\delta$ to be positively correlated to $\overline{\delta}$, which means a plot close to the main diagonal.
The AE and PCA variants show the closest plots to the main diagonal, thus, best stability. At the other extreme, TF-t-SNE shows widely varying 2D change for similar $n$D change, thus, high instability. Finally, Figure~\ref{fig:trails_cartolastd}e shows the absolute rank difference histograms for change. Their interpretation follows the one for the distance-preservation histograms (Fig.~\ref{fig:trails_cartolastd}c):
Left peaked histograms indicate high stability, whereas flatter ones indicate a discrepancy in 2D \emph{vs} $n$D changes. These histograms strengthen the insights obtained so far, making it even clearer that the AE and G-PCA methods are far stabler than the t-SNE, UMAP and TF-PCA.


\vspace{-0.15cm}
\section{Understanding dynamic projection behavior}
\label{sec:discussion}
The coarse-grained and fine-grained analyses presented so far highlighted that there are significant differences in the behavior of dynamic DR methods that depend on both the method and the dataset.
In this process, we also saw that visual quality and stability seem to be, in general, mutually competing for concerns -- methods that are good in one are not the best in the other.
We further explore these observations as follows. First, we analyze the causes of the observed (lack of) stability and link these to the way the studied DR techniques operate (Sec.~\ref{sec:unstable}). Next,
we summarize all our findings and propose a workflow to assist the practitioner in selecting a suitable DR technique for projecting dynamic data (Sec.~\ref{sec:choice}).

\begin{figure*}[!tb]
\centering
\includegraphics[width=0.745\linewidth]{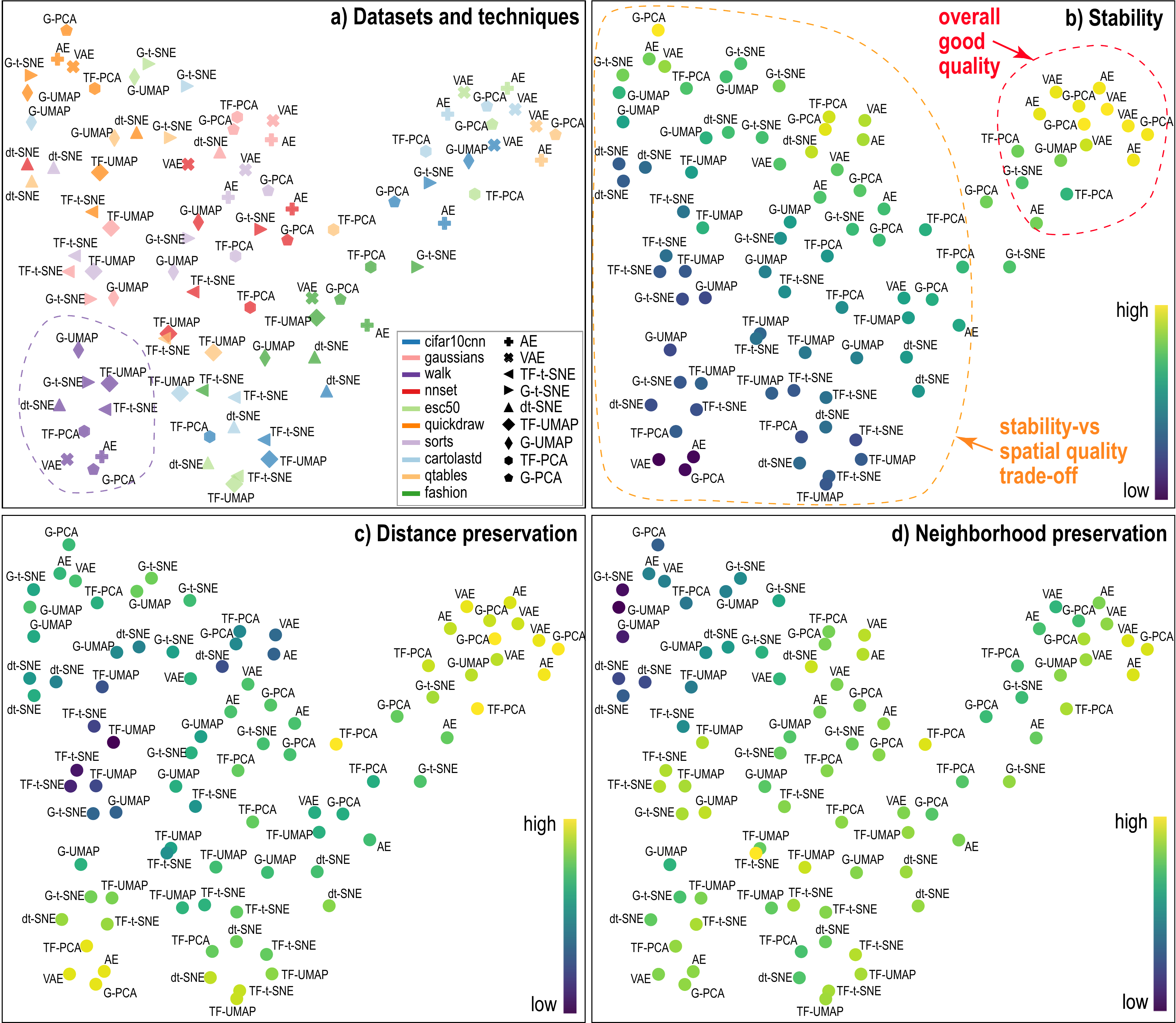}
\caption{Projection of projections map showing the similarity of all evaluated techniques on all datasets (Sec.~\ref{sec:choice}).}
\label{fig:tsne_0}
\vspace{-0.15cm}
\end{figure*}

\subsection{Analysis of (un)stable behavior}
\label{sec:unstable}
Beside empirically measuring and observing that different DR techniques have widely different stabilities, it is useful to analyze the \emph{causes} of these differences, which we do next.

\noindent\textbf{t-SNE and UMAP:} Our results tell that TF-t-SNE and TF-UMAP, that is, projections computed independently for each timestep, are the most unstable of the assessed techniques.
This is so since these are stochastic methods that optimize non-convex objective functions using randomly seeded gradient descent. Hence, different runs with the same data can create projections where different clusters might be formed and/or placed at different 2D positions. Figure~\ref{fig:instability}a,b shows the last scenario. From timesteps 1 to 2 of the TF-t-SNE run of the \textit{fashion} dataset, even though the local structure remains the same, the absolute position of the points and clusters changes drastically. In conclusion, using t-SNE/UMAP independently per timeframe is definitely not a good option for dynamic data.

\noindent\textbf{dt-SNE:} We encountered several cases where dt-SNE seems to have trouble optimizing its objective function -- for details, see the videos for \textit{qtables} and \textit{sorts}. In both these cases, dt-SNE did not capture any of the spatial structures present in the data, nor produced any sensible movement. These visual findings can be confirmed by the dark lines (low-quality values) in Fig. \ref{fig:all_datasets}. We also noticed that dt-SNE is very sensitive to the choice of hyperparameters. Concluding, whereas the initial findings in \cite{Rauber2016}, obtained on a single dataset (\emph{gaussians}) position this technique as a good option for projecting dynamic data, our additional findings raise questions about the practical value of this technique.

\noindent\textbf{PCA:} We also see instability in TF-PCA, but for different reasons than the ones discussed above. Specifically, if there is a change in rank of the top two eigenvectors from timestep $t$ to the next one, \emph{i.e.}, one of the associated eigenvalues becomes larger than the other, the projection exhibits an artifact that resembles a \emph{reflection} -- see the \emph{quickdraw} dataset in the two timesteps in Fig.~\ref{fig:instability}b,c. Alternatively, if the data changes sufficiently for the eigenvectors to change considerably, the projection shows a \emph{rotation}-like artifact -- see the two timesteps in Fig.~\ref{fig:instability}d,e. In contrast to t-SNE and UMAP, these artifacts are not due to stochastic seeding, but due to the way PCA works. Given the above, it is now clear why G-PCA is very stable -- it chooses the two largest-variation axes for the \emph{entire} dataset (all timesteps). The price to pay for this stability is that G-PCA may not yield the axes that best describe the data variation at each timestep, thus not the best spatial quality.

\noindent\textbf{Autoencoders:} Similarly to G-PCA, these techniques are stable since they train with the entire dataset (all timesteps) to learn a latent representation that encodes the global data distribution. Once trained, the encoder is a deterministic function that maps $nD$ data to 2D. The main disadvantage of autoencoders over G-PCA is usability: PCA is simple to implement and use. Autoencoders, in contrast, have the `usual' deep learning challenges, most notably finding the optimal network architecture and hyperparameter values.



\subsection{Finding similarly behaving techniques}
\label{sec:choice}
Figure~\ref{fig:aggregated} showed a high-level aggregated view of the quality metrics of the studied DR techniques, outlining that the autoencoders and PCA variants score better, in general, on both spatial quality and stability, than graph neighborhood techniques (t-SNE, dt-SNE, and UMAP). However, that image (and related analysis) was too aggregated. At the other extreme, Fig.~\ref{fig:all_datasets} and related discussion showed a fine-grained analysis of all metrics measured for all techniques run on all datasets. From both these analyses, it is quite hard to understand how (and when) different techniques behave similarly. This is arguably important for practitioners interested in choosing a technique in a given context (dataset type and metrics to maximize).

Figure~\ref{fig:tsne_0} supports this similarity analysis, as follows. Each point is here a technique run on a dataset, attributed by the computed 12 quality metrics. We project these points to 2D using UMAP, thus, creating a `projection of projections' map. The four images in Fig.~\ref{fig:tsne_0} use different visual codings to reveal several insights, as follows. Image (a) shows the techniques and datasets, coded by glyph, respectively categorical colors. Points in this plot are clustered more due to \emph{datasets} than \emph{techniques} -- that is, quality is more driven by the dataset nature than by which projection technique is used. For instance, we see the \emph{sorts} dataset well-separated as the purple cluster bottom-left in Fig.~\ref{fig:tsne_0}a. Images (b-d) show the same projection, colored by stability, distance preservation, and neighborhood preservation, respectively. The left part of the projection (orange dashed line, Fig.~\ref{fig:tsne_0}b) shows cases where stability and distance (and/or neighborhood) preservation are mutually complementary, \emph{i.e.}, when we obtain high stability, we get low distance/neighborhood preservation and conversely. The top-right part of the projection  (red dashed line, Fig.~\ref{fig:tsne_0}b) shows cases where both stability and spatial quality are quite high. All these cases use the AE, VAE, and G-PCA techniques. The central area of the projection is covered mainly by t-SNE, dt-SNE and UMAP, telling that these projections have average behavior (as compared to autoencoders and PCA variants). Looking at the color-coded plots (images b-d), we see that these projections do not score highest on any of the considered metrics.


The plots in Fig.~\ref{fig:tsne_0} can guide choosing a DR technique to project dynamic data: Given a dataset $\mathbf{D}$ to project, (1) find the most similar dataset $\mathbf{D}'$ in the benchmark, \emph{i.e.}, that contains data of similar nature (\emph{e.g.}, natural images, sounds) and is obtained via a similar acquisition process; (2) decide what is important for the dynamic projection of $\mathbf{D}$ -- stability, distance preservation, neighborhood preservation, or a mix of them; (3) find the projection techniques $P$ in the respective quality plots that have the desired qualities on $\mathbf{D}'$, and possibly also consider other projection techniques that behave similarly (close points in the plots). These techniques $P$ are then good candidates to project $\mathbf{D}$ with.

\section{Conclusion}
\label{sec:conclusion}
This paper is an initial step towards understanding the behavior of dimensionality reduction techniques in the context of dynamic/temporal data. We hope that the information and results presented here help practitioners who want to understand their complex data and that this work can be used by authors interested in developing DR techniques as a tool for evaluation and comparison.
We proposed a publicly available benchmark with 11 methods, ten datasets, and 12 quality metrics. To evaluate the viability of different techniques for the task, we computed spatial and temporal stability metrics for all possible combinations, thus providing an extensive collection of results. Based on the results, we presented a discussion that elaborates on the causes for understanding the dynamic behavior. All our experiments are documented and detailed online \cite{repo} to allow further analysis and reproducibility.

There are many ways this work can be extended in the future.
The benchmark can be extended with new methods, a better way to choose hyperparameters, new datasets, and new metrics. With a larger number of datasets, we can perform a robust test of the impact of dataset traits on the metrics.
We can also integrate streaming data techniques, datasets, and tests.

\section{Acknowledgements}
This study was financed in part by CAPES (Finance Code 001) and CNPq (Process 308851/2015-3).


\bibliographystyle{eg-alpha-doi}

\bibliography{paper}

\end{document}